\documentclass[12pt]{article}
\usepackage{amsmath,amsthm}
\usepackage{psfrag}
\usepackage{graphicx}
\usepackage{amssymb,latexsym}

\newtheorem{theorem}{Theorem}[section]
\newtheorem{lemma}[theorem]{Lemma}

\newtheorem{corollary}[theorem]{Corollary} 

\theoremstyle{remark}
\newtheorem{remark}[theorem]{Remark}

{\hspace*{\fill}$\rule{.3\baselineskip}{.35\baselineskip}$\end{trivlist}}

\newcommand{\bi}{\begin{itemize}}
\newcommand{\ei}{\end{itemize}}
\newcommand{\bt}{\begin{theorem}}
\newcommand{\et}{\end{theorem}}
\newcommand{\bp}{\begin{proof}}
\newcommand{\ep}{\end{proof}}
\newcommand{\be}{\begin{equation}}
\newcommand{\ee}{\end{equation}}
\newcommand{\ben}{\begin{enumerate}}
\newcommand{\een}{\end{enumerate}}

\newcommand{\C}{\mathbb C}

\newcommand{\N}{\mathbb N}

\newcommand{\R}{\mathbb R}

\newcommand{\Rscr}{\mathcal R}

\newcommand{\hf}{\frac{1}{2}}

\newcommand{\e}{\varepsilon}

\renewcommand{\O}{\Omega}

\renewcommand{\b}{\beta}

\newcommand{\z}{\zeta}

\renewcommand{\a}{\alpha}

\newcommand{\g}{\gamma}

\renewcommand{\and}{\text{~~ and ~~}}
\renewcommand{\part}{\partial}
\newcommand{\ra}{\rightarrow}

\newcommand{\gt}{\hat \gamma}

\begin{document}

\title{Determinant form of modulation equations for the semiclassical focusing  Nonlinear Schr\" odinger equation}

\author{Alexander Tovbis \footnote{
Department of Mathematics,
University of Central Florida, 
Orlando, FL 32816, email: atovbis@pegasus.cc.ucf.edu~~~Supported by  NSF grant DMS 0508779}  and
Stephanos Venakides
\footnote{
Department of Mathematics,
Duke University,
Durham, NC 27708, e-mail:
ven@math.duke.edu~~~
Supported by   NSF grant DMS
0207262}}

\begin{abstract}
We derive a determinant formula for the WKB exponential of singularly perturbed Zakharov-Shabat
system that corresponds to the semiclassical (zero dispersion) limit of the focusing  
Nonlinear Schr\" odinger equation. The derivation  is based on the Riemann-Hilbert Problem (RHP) representation  of the WKB exponential. We also 
prove its independence of the branchpoints of the corresponding hyperelliptic surface assuming
that the modulation
equations are satisfied.
\end{abstract}
\maketitle

\section{Introduction}

\begin{figure}
\centerline{
\includegraphics[height=6cm]{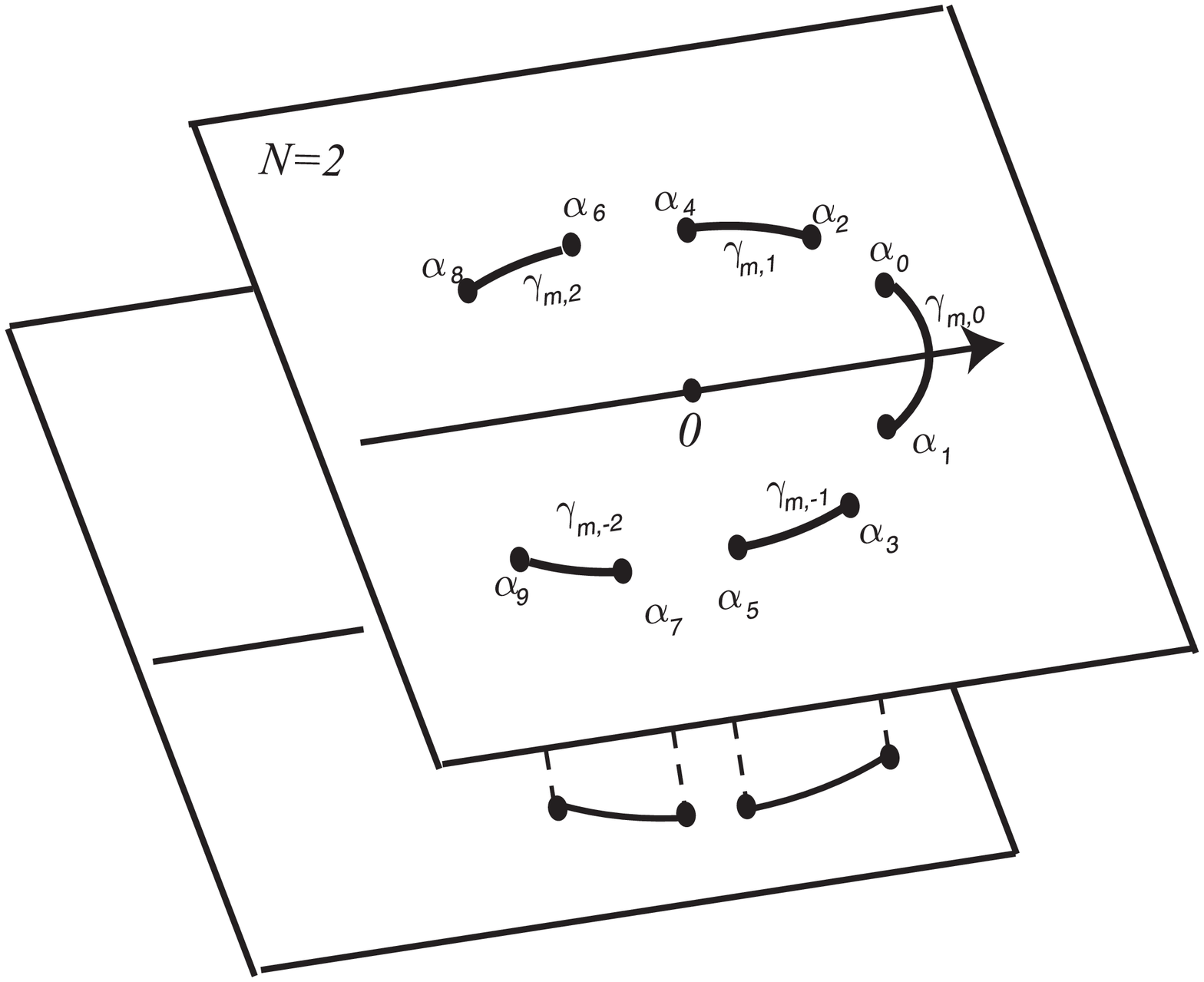}}
\caption{Riemann surface $\Rscr(x,t)$}
\label{hprell}
\end{figure}

The semiclassical analysis of the focusing Nonlinear
Schr\"odinger (NLS) 
\be  \label{FNLS}
i\e\partial_t q + \hf\e^2\partial_x^2 q + |q|^2q=0, \ \ \ \ \ \
\ee
has produced \cite{MK,KMM,TVZ1} solutions that one recognizes as modulated multi-phase periodic or quasi-periodic  waves. 
These wave solutions of NLS  are expressed  in terms of hyperelliptic  theta functions (see \cite{FT})
 built from the radical 
\be
R(z)=\left[\prod_{i=0}^{4N+1}(z-\a_i)\right]^{1/2},
\ee
where even $\a_{2k}$ lie in the upper halfplane and $\a_{2k+1}=\bar\a_{2k}$. 
Let $\Rscr=\Rscr(x,t)$ denote the hyperelliptic Riemann surface of $R$, where oriented arcs 
$\g^+_{m,k}$, connecting
$\a_{4k-2}$ and $\a_{4k}$, $k=1,2,\cdots,N$, their complex conjugates 
$\g^-_{m,k}$,
 connecting
$\a_{4k+1}$ and $\a_{4k-1}$, together with $\g_{m,0}$ connecting $\a_1$ and $\a_0$,
form the branchcuts (main arcs), see Fig. \ref{hprell}.

Points $\a_i$ depend
on $x,t$ but do not depend on $\e$. They are called
branchpoints of the hyperelliptic Riemann surface $\Rscr$ or simply branchpoints of $R$.
The number of wave-phases of a solution $q(x,t,\e)$ of \eqref{FNLS} (in the limit $\e\ra 0$) 
is equal to the {\it genus} $2N$ of the corresponding Riemann surface $\Rscr(x,t)$. 
The branchpoints satisfy  a system of equations 
 known as the {\it modulation equations} or the {\it modulation system} that is discussed below.

Nonlinear Schr\"odinger equation is one of the most celebrated examples of an integrable PDE,
i.e., a nonlinear PDE that can be ``linearized'' through the Lax pair. The ``$x$'' (spatial) part  
of the Lax pair is a second order system of linear ODEs
\be\label{ZS}
\partial_x \Phi=-\dfrac{i}{\e}
\begin{pmatrix}z&q(x,t,\e)\\ \bar q(x,t,\e) &-z\end{pmatrix}\Phi.
\ee
known as Zakharov-Shabat system (see 
\cite{ZS}). In the limit $\e\ra 0$
(semiclassical limit of the NLS) system \eqref{ZS} becomes a {\it singularly perturbed} system,
and, as such, is a subject of the WKB analysis. 

The goal of this paper is to find a determinant formula (see
\eqref{hK}) for the WKB exponential $g$ of \eqref{ZS}, to prove its independence of the branchpoints $\a$ assuming
that $\a$ satisfy the modulation
equations, and to derive various forms of modulation equations for $\a$ using the determinant
formula.  Our derivation  is based on the RHP representation \eqref{rhpg} of $g$, which was obtained
and discussed in \cite{TVZ1} for \eqref{FNLS} with pure radiational or 
 radiational and  solitons initial data
(for the pure soliton case, see \cite{KMM}). Connection between
the WKB exponential $g$ and the RHP \eqref{rhpg} is studied in a separate paper \cite{TV2}. The RHP
\eqref{rhpg} and its solution are the main objects of this paper.
 
The inverse scattering method of integration of \eqref{FNLS} is based on the scattering transform,
i.e., on the connection between the initial data of \eqref{FNLS} and the scattering data of 
\eqref{ZS}. In general, the scattering data consists of the reflection coefficient  defined on
$\R$ and of the eigenvalues of Zakharov-Shabat system together with their norming constants
(these eigenvalues correspond to solitons). In the semiclassical limit of \eqref{FNLS}, the reflection coefficient
$r_0(z,\e)$ depends on $z$ and $\e$. We denote 
\be\label{rf}
r(z,\e)=r_0(z,\e)e^{\frac{2i}{\e}(xz+2tz^2)}~~~~{\rm and}~~~~f=\hf \e i\ln r.
\ee
In general, $f$ is a function of $z$ and $\e$. However, studying the semiclassical 
limit of \eqref{FNLS}, we can consider only the leading order term (in $\e$) of $f$,
see \cite{TVZ1}. Throughout this paper, we
 assume that $f(z)$ has analytic continuation into the upper halfplane with the exception of a finite
number of logarithmic branchcuts and of isolated singularities. We further assume that the main 
arcs (branchcuts) of the hyperelliptic  surface $\Rscr(x,t)$  do not intersect the singularities of $f(z)$. The values of $f(z)$ in the lower halfplane
are obtained by Schwarz reflection. Thus, in general, $\Im f(z)$ has a jump on the real axis.

To define modulation equations, we first assume that the branchpoints $\a_{2j}=\a_{2j}(x,t)$, 
$j=0,1,\cdots,2N$, are known.  Let  $\g^\pm_{c,k}$ be  oriented arcs connecting $\a_{4k-4}, 
\a_{4k-2}$  and $\a_{4k-1},\a_{4k-3}$ respectively. These arcs are called {\it complementary arcs}.
Let $\g_{m,k}=\g^+_{m,k}\cup\g^-_{m,k}$, $\g_{c,k}=\g^+_{c,k}\cup\g^-_{c,k}$, $k=1,2,\cdots, N$, where
orientation is inherited.
Define $g(z)$ as the solution of
the RHP:
\begin{align}\label{rhpg}
g_++g_-&=f+W_j ~~\mbox{ on the main arc $\g_{m,j}$, $j=0,1,\cdots, N$} \cr
g_+ - g_-&=\O_j~~\mbox{ on the complementary arc $\g_{c,j}$, $j=1,\cdots, N$ }\cr
g(z) &  ~~\mbox{is analytic at}~~ z=\infty~,
\end{align}
where all $W_j,\O_j$ are some real constants. We assume that the real
constant $W_0=0$ on the main arc $\g_{m,0}$ that connects $\a_1$ and $\a_0$. 

\begin{figure}
\centerline{
\includegraphics[height=6cm]{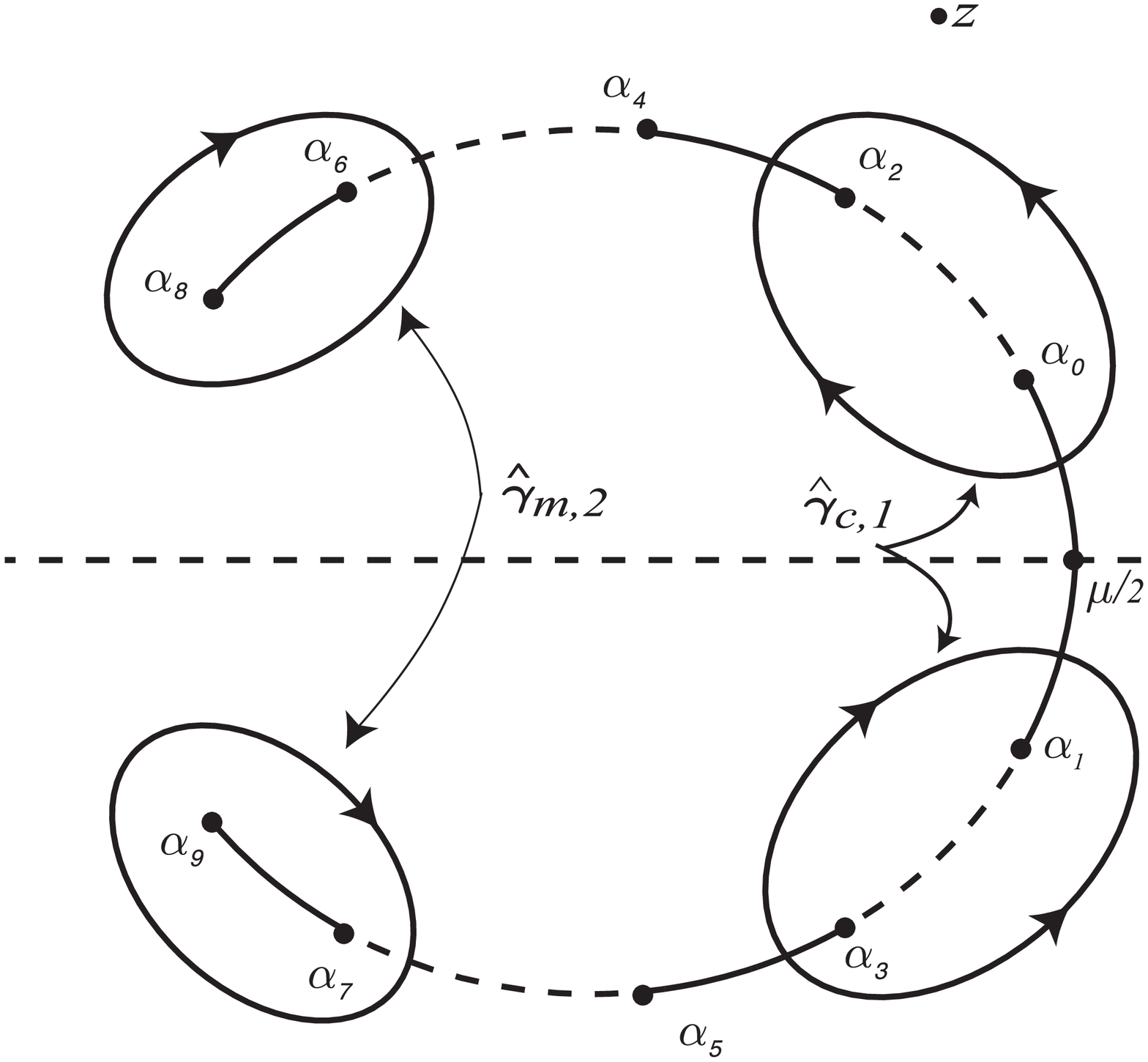}}
\caption{Contours  $\gt_{m,2},~\gt_{c,1}$ }
\label{figloop}
\end{figure}

Let $\g$ denote the unioun of all the main arcs $\g_{m,j}$, $j=0,1,\cdots, N$ and all the 
complementary arcs with the inhereted orientation.
It is well known that solution 
to the RHP \eqref{rhpg} is given by 
\begin{equation}\label{gform1}
g(z)={{R(z)}\over{2\pi i}}\left[ \int_\g{{f(\z)}\over{(\z-z)R(\z)_+}}d\z
+\sum_{i=1}^N\left( \int_{\g_{m,i}}{{W_i}\over{(\z-z)R(\z)_+}}d\z
+\int_{\g_{c,i}}{{\Omega_i}\over{(\z-z)R(\z)}}\right) d\z\right]~.  
\end{equation}
Expressing the integrals over the arcs as integrals over the loops shown in Fig. \ref{figloop}, 
we obtain 
\begin{equation}\label{gform2}
g(z)={{R(z)}\over{4\pi i}}\left[ \oint_{\gt}{{f(\z)}\over{(\z-z)R(\z)}}d\z
+\sum_{i=1}^N\left( \oint_{\gt_{m,i}}{{W_i}\over{(\z-z)R(\z)}}d\z
+\oint_{\gt_{c,i}}{{\Omega_i}\over{(\z-z)R(\z)}}\right) d\z\right],  
\end{equation}
where  the loops $\gt$ and  $\gt_{m,i}$ and the contours $\gt_{c,i}$ ($\gt_{c,i}$ 
consists of the sum of two arcs oriented oppositely as in the figure) are  
contractible  to their corresponding arcs without passing through $z$.  

Deforming $\gt$ so that now $z$ {\it is inside the loop} $\g$ and still outside the loops $\gt_{m,i}$ and 
$\gt_{c,i}$, we obtain 
\begin{equation}\label{hform}
h(z)={{R(z)}\over{2\pi i}}\left[ \oint_{\gt}{{f(\z)}\over{(\z-z)R(\z)}}d\z
+\sum_{i=1}^N\left( \oint_{\gt_{m,i}}{{W_i}\over{(\z-z)R(\z)}}d\z
+\oint_{\gt_{c,i}}{{\Omega_i}\over{(\z-z)R(\z)}}\right) d\z\right]. 
\end{equation}
where 
\be h(z)=2g(z)-f(z).\ee  
The function $h(z)$ is obtained by multiplying $g$ by a factor of $2$ and the 
residue $-f$ being picked up as $z$  cuts through the loop $\gt$. 

If $z$ approaches the $i$th main arc from either side, a residue is generated 
as $z$ cuts through the loop $\gamma_{m,i}$ encircling the arc; multiplied by the factor 
${{R(z)}\over{2\pi i}}$ outside the integral,  the residue yields the contribution 
$W_i$ to $h$. Similarly, if $z$ approaches the $i$th complementary arc, the contribution 
to $h$ from $z$ cutting the contour $\gamma_{c,i}$ is $+\O_i$ or $-\O_i$, depending 
on whether $z$ approaches from the left or right of the contour. These observations 
lead directly to the jump conditions
\begin{align}\label{rhph}
h_++h_-&=2W_j ~~\mbox{ on the main arc $\g_{m,j}$, $j=0,1,\cdots, N$ } \cr
h_+ - h_-&=2\O_j~~\mbox{ on the complementary arc $\g_{c,j}$, $j=1,\cdots, N$  }\cr
h_+- h_- &= -2i\Im f  ~~\mbox{on the real axis.}
\end{align}
To see this, one takes into account the above 
residue calculations and the fact that the expression in the square brackets in \eqref{hform} is analytic 
in a neighborhood of any $\a_j$ (we are assuming distinct $\a_j$).  
From the above, it is clear that at any $\a=\a_{2j}$
\be\label{hasymp0}
h(z)\sim W+\pm\O+\nu_1(z-\a)^{\frac{1}{2}}+\nu_3(z-\a)^{\frac{3}{2}}+\cdots. 
\ee
where $\pm$ refers to whether $z$ is left or right of the contour and $W,\O$
denote real constants on the main  and the complementary arcs, adjacent to $\a=\a_{2j}$.

According to \eqref{gform2}, $g(z)\sim O(z^{2N})$ as $z\ra\infty$. 
The requirement that $g(z)$ is analytic at $z=\infty$ defines the system of $2N$ 
linear equations for $W_j,\O_j$
\begin{equation}\label{moments}
\oint_{\gt}{\z^k {f(\z)}\over{R(\z)}}d\z
+\sum_{i=1}^N\left( \oint_{\gt_{m,i}}{{W_i\z^k}\over{R(\z)}}d\z
+\oint_{\gt_{c,i}}{{\Omega_i\z^k}\over{R(\z)}}d\z\right)=0,  
 \ \ \ \ \
k=0,1,\cdots, 2N-1.
\end{equation}

Modulation equations comes from the requirement that the $L^2$-solution to the RHP
\eqref{rhpg}, i.e., the Cauchy operator in \eqref{gform1}, commutes with the differentiation in $z$. The equivalent
statement is that at every $\a=\a_{2j}$ the coefficient $\nu_1=0$, so that \eqref{hasymp0}
becomes
\be\label{hasymp}
h(z)\sim W_j\pm\O_j+\nu_3(z-\a)^{\frac{3}{2}}+\cdots.  
\ \ \ \mbox{as} \ z\to\a.
\ee
Equation \eqref{hasymp} implies that the expression in the square brackets in equation 
\eqref{hform}, let us call it $B(z)$, 
vanishes at every $\a_{2j}$, i.e. 
\be\label{modeq1}
B(\a_{2j})=\oint_{\gt}{{f(\z)}\over{(\z-\a_{2j})R(\z)}}d\z
+\sum_{i=1}^N\left( \oint_{\gt_{m,i}}{{W_i}\over{(\z-\a_{2j})R(\z)}}d\z
+\oint_{\gt_{c,i}}{{\Omega_i}\over{(\z-\a_{2j})R(\z)}}\right) d\z=0, 
\ee
where $\ j=0, 1, \cdots, 2N$ and $\a_{2j}$ is {\it inside} the loops around the main and the
complementary arcs
 that are adjacent to $\a_{2j}$ but outsisde all other loops 
$\gt_{m,i}$ and $\gt_{c,k}$.
The {\it modulation  equations} \eqref{modeq1} is a system of $2N+1$ complex  
conditions satisfied by the $2N+1$ complex branchpoints  $\a_{2j}$, $j=0,1,\cdots, 2N$. It is also
satisfied   by
their complex conjugates $\a_{2j+1}$. 

\section{Determinant formula}

To simplify the notations, we consider below the case $N=1$. The obtained 
formulae allow a straightforward generalization to the arbitrary $N\in \N$ case.
Let 
\be\label{D,K}
D=\left|\begin{matrix}\oint_{\gt_m}\frac{ d\z}{R(\z)} &\oint_{\gt_m}\frac{\z d\z}{R(\z)} \cr
\oint_{\gt_c}\frac{ d\z}{R(\z)} & \oint_{\gt_c}\frac{\z d\z}{R(\z)}\end{matrix}\right|,~~~~~
K(z)= \frac{1}{2\pi i}\left| \begin{matrix} 
\oint_{\gt_m}\frac{ d\z}{R(\z)} &\oint_{\gt_m}\frac{\z d\z}{R(\z)}
& \oint_{\gt_m}\frac{d\z}{(\z-z)R(\z)} \\ 
\oint_{\gt_c}\frac{ d\z}{R(\z)} & \oint_{\gt_c}\frac{\z d\z}{R(\z)} & \oint_{\gt_m}\frac{d\z}{(\z-z)R(\z)} \cr
\oint_{\gt}\frac{f(\z)d\z}{R(\z)} &  \oint_{\gt}\frac{\z f(\z)d\z}{R(\z)} &
\oint_{\gt}\frac{f(\z)d\z}{(\z-z)R(\z)}\cr
\end{matrix}\right|,
\ee
where $\g_m=\g_{m,1}$ and $\g_c=\g_{c,1}$.
It is well known that $D\ne 0$ if all $\a_{2j},~j=0,1,2$ are distinct. 

Multiplying the first two rows of $K(z)$ by $W=W-1$ and $\O=\O_1$ respectively, adding them
to the third row and utilizing \eqref{hform} and \eqref{moments}, we obtain
\be\label{hK}
h(z)=\frac{R(z)}{D}K(z)
\ee
where $z$ is inside the loop $\gt$ but outside the loops $\gt_m,\gt_c$.
That will be our standard assumption about the location of $z$ for the rest of the
paper, unless specified otherwise. 
It is clear that moving $z$ inside the loops $\gt_m,\gt_c$ would  generate residue terms $W$ and $\pm\O$ (depending on the direction $z$ crosses the oriented loop $\gt_c$) in the right hand side of \eqref{hK}.
Combining this fact with \eqref{hasymp}, we obtain a new form of modulation equations
\be\label{Kmod}
K(\a_{2j})=0,~~~~~~~j=0,1,2.
\ee

\begin{lemma}\label{derKa}
Let $\a$ denote one of the branchpoints. Then
\be\label{dK/da}
\frac{\part K(z)}{\part \a}=\frac{h(z)}{R(z)}\left[ \frac{D}{2(z-\a)} + 
\frac{\part D}{\part \a}\right]~.
\ee 
\end{lemma}

\bp
Let us write  $K(z)=\frac{1}{2\pi i}\left| K_1, K_2, K_3(z)\right|$, where $K_j$ denote 
the $j$th column of
the determinant $K(z)$, see \eqref{D,K}. Using
\be\label{dR/da}
\frac{\part}{\part \a}\frac{1}{ R(\z)}=\frac{1}{2(\z-\a) R(\z)}~~~~{\rm and}~~~~ 
\frac{\z}{\z-\a}=1+\frac{\a}{\z-\a}~,
\ee
we obtain
\be\label{dK12/da0}
\left|\frac{\part K_1}{\part \a}, K_2, K_3(z)\right|=\hf\left| K_3(\a), K_2, K_3(z)\right|,~~~~~
\left|K_1, \frac{\part K_2}{\part \a},  K_3(z)\right|=\hf\left|K_1, \a K_3(\a),  K_3(z)\right|~.
\ee
Multiplying the first two rows of each of the above two determinants  by $W$ and $\O$ respectively, adding them
to the third row and utilizing \eqref{hform}, \eqref{moments} and
\eqref{Kmod} with $\a_{2j}=\a$, we obtain (as before)
\begin{align}\label{dK12/da1}
\frac{1}{2\pi i}\left|\frac{\part K_1}{\part \a}, K_2, K_3(z)\right|=&
\hf\left|\begin{matrix}\oint_{\gt_m}\frac{ d\z}{(\z-\a)R(\z)} &
\oint_{\gt_m}\frac{\z d\z}{R(\z)} \cr
\oint_{\gt_c}\frac{ d\z}{(\z-\a)R(\z)} & \oint_{\gt_c}\frac{\z d\z}{R(\z)}\end{matrix}\right|
\frac{h(z)}{R(z)}, \cr
\frac{1}{2\pi i}\left|K_1, \frac{\part K_2}{\part \a},  K_3(z)\right|=&
\hf\left|\begin{matrix}\oint_{\gt_m}\frac{ d\z}{R(\z)} 
&\oint_{\gt_m}\frac{\z d\z}{(\z-\a)R(\z)} \cr
\oint_{\gt_c}\frac{ d\z}{R(\z)} & \oint_{\gt_c}\frac{\z d\z}{(\z-\a)R(\z)}\end{matrix}\right|
\frac{h(z)}{R(z)}. \cr
\end{align}
Adding these two determinants while  taking into account \eqref{D,K} yields
\be\label{dK1+dK2}
\frac{1}{2\pi i}\left[ \left|\frac{\part K_1}{\part \a}, K_2, K_3(z)\right|+
\left|K_1, \frac{\part K_2}{\part \a},  K_3(z)\right|\right] = \frac{\part D}{\part \a} \cdot \frac{h(z)}{R(z)}~.
\ee
Notice that 
\be\label{dK3/da}
\frac{\part K_3(z)}{\part \a}=\frac{1}{2(z-\a)}\left[K_3(z)-K_3(\a) \right] 
\ee
follows from
$\frac{1}{(\z-\a)(\z-z)}=\frac{1}{z-\a}\left[\frac{1}{\z-z} - \frac{1}{\z-\a} \right]$ and
\eqref{dR/da}. Using again \eqref{Kmod} with $\a_{2j}=\a$, we obtain
\be\label{dK3}
\frac{1}{2\pi i}\left|K_1, K_2, \frac{\part K_3(z)}{\part \a}\right| = \frac{1}{2(z-\a)}K(z)~.
\ee
Adding \eqref{dK1+dK2} and \eqref{dK3} completes the proof.
\ep

Lemma \ref{derKa} is the basis for the following theorem.

\begin{theorem}\label{dhda=0}
Let $\a$ denote one of the branchpoints. Then the equation $K(\a)=0$ implies
\be\label{dh/da}
\frac{\part }{\part \a}h(z)\equiv 0~.
\ee 
\end{theorem}

\bp
According to \eqref{hK}, we have 
\be\label{dK/dah}
\frac{\part K(z)}{\part \a}=\frac{\part h(z)}{\part \a}\cdot\frac{D}{R(z)}+
h(z)\frac{\part }{\part \a}\frac{D}{R(z)}~.
\ee
Substituting \eqref{dK/da} and \eqref{dR/da} into \eqref{dK/dah}, we obtain
\be
\frac{D}{R(z)}\frac{\part }{\part \a}h(z)\equiv 0~,
\ee
which implies \eqref{dh/da}.
\ep

\begin{corollary}\label{dhdxt=0}
Modulation equations $K(\a_{2j})=0$, $j=0,1,2$ imply
\be\label{dh/dxt}
\frac{d}{dx}h(z)\equiv \frac{\part }{\part x}h(z),~~~~~~~
\frac{d}{dt}h(z)\equiv \frac{\part }{\part t}h(z)~.
\ee 
\end{corollary}

\begin{remark}\label{arbitrN}
All the results of this section, unless mentioned otherwise, remain true for arbitrary 
genus $2N,~N\in \N$. Moreover, they  do not depend on Schwarz symmetry of $\g$ and of $f(z)$, as well
as on any particular form of the functional dependence of $f(z)=f(z;\b)$ on the external parameter(s)
$\b$ (in the  discussion above, $\b=x,t$) and, in fact, are true in much more general setting.
Indeed, let $\g=\g(\b)$ be a Jordan curve in $\C$ and let 
$f(z)=f(z;\b)$ be analytic (in $z$) on some
open set $S\supset\g$ and smooth in $\b$. The contour $\g$ is partitioned into a finite number
of interlaced nondegenerate (positive measure) main and complementary arcs by the branchpoints
$\a_j\in\g$, $j=1,2,\cdots, 2n$. Then we have $n$ main arcs $\g_{m,j}$ and $n$ or $n-1$,
depending on whether $\g$ is closed or not, complementary arcs $\g_{m,j}$. The genus of the
hyperelliptic Riemann surface $\Rscr(\b)$ of the radical $R(z)=\sqrt{\prod_{j=1}^{2n}(z-\a_j)}$
is $n-1$. Let $g(z)=g(z;\b)$ satisfies the conditions \eqref{rhpg}, where $W_j$, $\O_j$ are some 
complex constants. In fact, all except any $N-1$ of these constants can be choosen arbitrarily.
Let $h(z)=2g(z)-f(z)$. Then the modulation equation \eqref{hasymp} at any $\a=\a_j$
implies $\frac{\part h(z)}{\part \a}\equiv 0$. 
\end{remark}

Since $D$ and $R$ do not explicitly depend on $x,t$, Corollary \ref{dhdxt=0} together with
\eqref{hK} imply that
\be\label{dhdxtdK}
\frac{d}{dx}h(z)=\frac{R(z)}{D}\frac{\part}{\part x} K(z),~~~
\frac{d}{dt}h(z)=\frac{R(z)}{D}\frac{\part}{\part t} K(z)~.
\ee
Using \eqref{D,K} and
\be
f(z)=f_0(z)-xz-2tz^2~,
\ee
where $f_0(z)=\hf \e i\ln r_0(z)$, we calculate
\be\label{dKdx} 
\frac{\part}{\part x} K(z)=
-\left|\begin{matrix}\oint_{\gt_m}\frac{ d\z}{(\z-z)R(\z)} &\oint_{\gt_m}\frac{ d\z}{R(\z)} \cr
\oint_{\gt_c}\frac{ d\z}{(\z-z)R(\z)} & \oint_{\gt_c}\frac{ d\z}{R(\z)}\end{matrix}\right|~,
\ee
and
\be\label{dKdt} 
\frac{\part}{\part t} K(z)=
2\left|\begin{matrix}\oint_{\gt_m}\frac{ d\z}{(\z-z)R(\z)}
&\oint_{\gt_m}\frac{\z d\z}{R(\z)} \cr
\oint_{\gt_c}\frac{ d\z}{(\z-z)R(\z)} & \oint_{\gt_c}\frac{\z d\z}{R(\z)}\end{matrix}\right|
+\sum_{j=0}^5 \a_j\frac{\part}{\part x} K(z)=
2\left|\begin{matrix}\oint_{\gt_m}\frac{ d\z}{(\z-z)R(\z)} &\oint_{\gt_m}\frac{\z-\hf \sum_{j=0}^5 \a_j d\z}{R(\z)} \cr
\oint_{\gt_c}\frac{ d\z}{(\z-z)R(\z)} & \oint_{\gt_c}\frac{\z-\hf \sum_{j=0}^5 \a_j d\z}{R(\z)}\end{matrix}\right|~,
\ee
where $z$ is outside the loops $\gt_m,\gt_c$.

It is easy to see that if $z$ is inside  loops $\gt_m,\gt_c$ the  $\gt_m$ integrals in
\eqref{dKdx}, \eqref{dKdt} are equal to the corresponding integrals on the segment $[\bar \a_0, \a_0]$
multiplied by $-2$, and $\gt_c$ integrals in
\eqref{dKdx}, \eqref{dKdt} are equal to the corresponding integrals on the segment $[\bar \a_4, \a_4]$
multiplied by $2$. If $z$ is outside any of the loops $\gt_m,\gt_c$, then the corresponding residues should be taken into account. 


\section{Differential form of modulation equations}

Modulation equations \eqref{Kmod} can be rewritten as ODEs
\be\label{dK/daa/xt}
\frac{\part K}{\part \a}\a_x=-\frac{\part}{\part x} K,~~~~~~~~~~~
\frac{\part K}{\part \a}\a_t=-\frac{\part}{\part t} K~,
\ee
where $\a$ denotes the vector $\a=(\a_0,\a_2,\a_4)$ (alternatively, we can consider 
$\a$ to be a 6-dimensional vector). According to \eqref{dK/da}, in the case $j\ne l$, $j,l=0,2,4$,
we have
\be\label{dkam/daj}
\frac{\part K(\a_l)}{\part \a_j}=\lim_{z\ra\a_l}\frac{h(z)}{R(z)}\left[ \frac{D}{2(z-\a_j)} + 
\frac{\part D}{\part \a_j}\right]~,
\ee
where $z$ is inside any of the loops $\gt_m,\gt_c$ that surround $\a_m$.  Then
$\lim_{z\ra\a_l}\frac{h(z)}{R(z)}=0$. So, $\frac{\part K(\a_l)}{\part \a_j}=0$.
That means that the matrix $\frac{\part K}{\part \a}$ is diagonal.

To calculate $\frac{\part K(\a_j)}{\part \a_j}$, we notice that 
\be\label{dK1a+dK2a}
\frac{1}{2\pi i}\left[ \left|\frac{\part K_1}{\part \a_j}, K_2, K_3(\a_j)\right|+
\left|K_1, \frac{\part K_2}{\part \a_j},  K_3(\a_j)\right|\right] = \frac{\part D}{\part \a_j} \cdot \lim_{z\ra\a_j}\frac{h(z)}{R(z)}=0~.
\ee
Let us take $j=2$. Then
\begin{align}
\label{dK3/daj} 
&\frac{\part K(\a_2)}{\part \a_2}=\frac{1}{2\pi i} \left|K_1, K_2, 
\frac{\part K_3(\a_2)}{\part \a_j}\right|=\cr
\frac{3D}{4\pi i}
&\left[ \oint_{\gt}{{f(\z)}\over{(\z-\a_2)^2R(\z)}}d\z
+\left( \oint_{\gt_{m}}{{W}\over{(\z-\a_2)^2R(\z)}}d\z
+\oint_{\gt_{c}}{{\Omega}\over{(\z-\a_2)^2R(\z)}}\right) d\z \right].
\end{align}
Equation \eqref{hasymp} in a vicinity of $z=\a_2$ can be rewritten as  
\be\label{hasymaj}
h(z)= W \pm\O +c_2(z-\a_2)R(z)+O(z-\a_2)^{5\over 2}
\ee
where $c_2\in \C$. If $z$ is inside any loops $\gt_m,\gt_c$, 
the constants $ W,\O$ in \eqref{hasymaj} should be replaced by zeroes, 
so we have 
\be\label{h/R}
\frac{h(z)}{R(z)}=c_2(z-\a_2)+O(z-\a_2)^2.
\ee
Differentiating \eqref{h/R} and taking into the account \eqref{hform}, we obtain
\begin{align}
\label{dKaj/daj}
&c_2=\lim_{z\ra\a_2}\left( \frac{h(z)}{R(z)}\right)'= \cr
\frac{1}{2\pi i}&
\left[ \oint_{\gt}{{f(\z)}\over{(\z-\a_2)^2R(\z)}}d\z
+\left( \oint_{\gt_{m}}{{W}\over{(\z-\a_2)^2R(\z)}}d\z
+\oint_{\gt_{c}}{{\Omega}\over{(\z-\a_2)^2R(\z)}}\right) d\z\right].
\end{align}
Thus,
\be\label{dK/daj}
\frac{\part K(\a_2)}{\part \a_2}=\frac{3}{2} c_2D~.
\ee

Formulae \eqref{dK3/daj}-\eqref{dK/daj} are also applicable for $\a_0$, $\a_4$,
if formulae \eqref{dK3/daj}, \eqref{hasymaj}, \eqref{dKaj/daj} contain only integrals
over the loops that are adjasent to $\a_0$, $\a_4$ respectively, and only the constants
$W,\O$ that correspond to these loops.

According to \eqref{dK3/daj}-\eqref{dK/daj}, \eqref{dK/daa/xt} can be written as
\be\label{axtodes}
\frac{3}{2}c_jD(\a_j)_x=-\frac{\part}{\part x} K(\a_j),~~~~~~~~~
\frac{3}{2} c_jD(\a_j)_t=-\frac{\part}{\part t} K(\a_j)~.
\ee
Ordinary differential equations \eqref{axtodes} imply
\be\label{axtpde}
(\a_j)_t=\dfrac{\frac{\part}{\part t} K(\a_j)}{\frac{\part}{\part x} K(\a_j)}(\a_j)_x,
\ee
where, according to \eqref{dKdx}, \eqref{dKdt}
\be\label{K/K}
\dfrac{\frac{\part}{\part t} K(\a_j)}{\frac{\part}{\part x} K(\a_j)}=
\sum_{j=0}^5 \a_j+2\frac{\left|\begin{matrix}\oint_{\gt_m}\frac{ d\z}{(\z-\a_j)R(\z)}
&\oint_{\gt_m}\frac{\z d\z}{R(\z)} \cr
\oint_{\gt_c}\frac{ d\z}{)\z-\a_j)R(\z)} & \oint_{\gt_c}\frac{\z d\z}{R(\z)}\end{matrix}\right|}
{\left|\begin{matrix}\oint_{\gt_m}\frac{ d\z}{(\z-\a_j)R(\z)} &\oint_{\gt_m}\frac{ d\z}{R(\z)} \cr
\oint_{\gt_c}\frac{ d\z}{)\z-\a_j)R(\z)} & \oint_{\gt_c}\frac{ d\z}{R(\z)}\end{matrix}\right|}.
\ee
This is the Riemann invariant form of modualtion equations written as PDEs.

Alternatively, diffrential form of the modulation equations \eqref{axtodes} can be obtained
by differentiating \eqref{hasymaj} and the corresponding equations at $\a_0,\a_2$. At $\a_2$ we have
\begin{align}
\label{dh/dxtloc}
\frac{d}{dx}h(z)&=(W)_x\pm(\O)_x-\frac{3}{2}c_jR(z)(\a_j)_x+O(z-\a_2),\cr
\frac{d}{dt}h(z)&=(W)_t\pm(\O)_t-\frac{3}{2}c_jR(z)(\a_j)_t+O(z-\a_2).
\end{align}
Moving $z$ inside 
any loops $\gt_m,\gt_c$ that surround $\a_2$ will eliminate derivatives of $W,\O$
in \eqref{dh/dxtloc}. According to \eqref{dhdxtdK}, \eqref{dKdx} and \eqref{dKdt}, these derivatives are
\be\label{const_x}
W_x=\frac{2\pi i}{D} \oint_{\gt_c}\frac{ d\z}{R(\z)},~~~~~~~~~~~~~
\O_x=-\frac{2\pi i}{D} \oint_{\gt_m}\frac{ d\z}{R(\z)}
\ee
and 
\be\label{const_t}
W_t=-\frac{4\pi i}{D} \oint_{\gt_c}\frac{\z-\hf \sum_{j=0}^5 \a_j }{R(\z)}d\z,~~~~~~~~~~~
\O_t=\frac{4\pi i}{D} \oint_{\gt_m}\frac{\z-\hf \sum_{j=0}^5 \a_j }{R(\z)}d\z.
\ee
Now equations \eqref{axtodes} follows from \eqref{dh/dxtloc} and \eqref{dhdxtdK}. Equations 
\eqref{const_x}-\eqref{const_t} also imply
\be\label{derivmatr}
\left|\begin{matrix}\O_x
& \O_t \cr
W_x & W_t\end{matrix}\right|=-\frac{8\pi^2}{D}~.
\ee

Finally, since the Cauchy operator for the RHP \eqref{rhpg} commutes with differentiation,
we have
\be\label{h'}
h'(z)={{R(z)}\over{2\pi i}} \oint_{\gt}{{f'(\z)}\over{(\z-z)R(\z)}}d\z~.
\ee
Combining this with \eqref{hasymaj} yield 
$h'(z)=\left[ \frac{3}{2}c_j + O(\sqrt{z-\a_j}\right]R(z)$ in a vicinity of $z=\a_j$. 
Thus,
\be\label{cj}
c_j={1\over{3\pi i}} \oint_{\gt}{{f'(\z)}\over{(\z-\a_j)R(\z)}}d\z~.
\ee 
Substitution of \eqref{cj} into \eqref{axtodes} yields
\be
(\a_j)_x=-\dfrac{2\pi i\frac{\part}{\part x} K(\a_j)}{D\oint_{\gt}{{f'(\z)}\over{(\z-\a_j)R(\z)}}d\z},~~~~~~~~~
(\a_j)_t=-\dfrac{2\pi i\frac{\part}{\part t} K(\a_j)}{D\oint_{\gt}{{f'(\z)}\over{(\z-\a_j)R(\z)}}d\z}~,
\ee
$j=0,2,4$.

\bibliographystyle{plain}
\bibliography{spin.bib}
\end{document}